\documentclass{ws-procs975x65}
\newcommand {\be}{\begin{equation}}
\newcommand {\ee}{\end{equation}}

\begin{document}

\title{INHOMOGENEOUS UNIVERSE IN LOOP QUANTUM GRAVITY}

\author{FRANCESCO CIANFRANI$^*$}

\address{Instytut Fizyki Teoretycznej, Uniwersytet Wroc\l awski, \\ 
pl. M. Borna 9, 50-204 Wroc\l aw, Poland, EU.\\
$^*$E-mail: francesco.cianfrani@ift.uni.wroc.pl}



\begin{abstract}
It is discussed a truncation of the kinematical Hilbert space of Loop Quantum Gravity, which describes the dynamical system associated with an inhomogeneous cosmological model. 
\end{abstract}


\bodymatter

\section*{}
The cosmological implementation of Loop Quantum Gravity (LQG) has been realized on a canonical level via Loop Quantum Cosmology (LQC)\cite{lqc}. This framework is based on first reducing the phase space according with the symmetries of the homogeneous Bianchi models and then implementing a polymer-like quantization of the resulting dynamical system. The connection with LQG is in the choice of reduced variables, which describe the dynamical part of the Ashtekar connections and the associated momenta when restricted to Bianchi models, and in the formal analogies between polymer and loop quantization. However, the classical reduction implies that the kinematical constraints are missing on a quantum level. In particular, the absence of the vector constraint is responsible for the issue of the scalar constraint regularization in LQC. In this proceeding and in the companion one \cite{ema}, we will define a formulation for a quantum cosmological space-time in which diffeomorphisms invariance is not completely broken and the regularization issue is solved. 

\paragraph{Inhomogeneous Bianchi I model} Let us consider an inhomogeneous extension of the Bianchi I model, in which reduced variables retain a dependence on spatial coordinates, {\it i.e.}\footnote{in what follows repeated internal indexes ($i,k,..$) will not be summed.}
\be
A^i_a=c_i(t,x)\delta^i_a,\quad E^a_i=p^i(t,x)\delta^a_i.\label{in}
\ee
These phase space variables can be obtained in two relevant case: i) by assuming that each scale factors $a_i$ is a function only of the associated Cartesian coordinate $x^i=\delta^i_a x^a$, ii) when spatial gradients of the scale factors are negligible with respect to their time derivatives. The case i) describes a reparametrized homogeneous Bianchi I model, for which it is possible to avoid the spatial dependence by a change of variables. The case ii) is appropriate to extend the Belinski Khalatnikov Lifshitz conjecture \cite{bkl} to loop variables and it can give an insight into the behavior of the generic inhomogeneous cosmological solution approaching the singularity.

Once the conditions (\ref{in}) hold, the kinematical constraints are not identically solved as in the homogeneous case\cite{art}. Instead, the Gauss constraint reduces to three independent conditions $G_i=0$. Given an integral curve $\Gamma_i$ of the vector $\partial_i=\delta_i^a\partial_a$, $G_i=0$ is \emph{the Gauss constraint associated with a U(1) gauge theory on $\Gamma_i$}, whose associated U(1) connection is $c_i$. We denote by $U(1)_i$ the three kinds of U(1) gauge symmetries we get in this way.
As soon as the vector constraint is concerned, it now generates what we call \emph{reduced diffeomorphisms}, {\it i.e.} the composition of a generic diffeomorphisms along a direction $i$ and rigid translations along the directions $k\neq i$. These reduced diffeomorphisms send a given integral curve $\Gamma_i$ into another one $\Gamma_i'$, which is still an integral curve of the same vector field $\partial_i$. 

\paragraph{Reduced Quantization} The quantization can be carried on by quantizing the algebra of the holonomies along $\Gamma_i$ for $i=1,2,3$ and of the fluxes across the associated dual surfaces. What it turns out \cite{art} is that the kinematical Hilbert space can be defined as 
\be
\mathcal{H}=\bigotimes_{i=1}^3\mathcal{H}_i,\qquad \mathcal{H}_i=\mathcal{L}^2(U(1)_i,d\mu_i),
\ee    
where $\mathcal{H}_i$ is the space of the square integrable $U(1)_i$ functionals over all the possible $\Gamma_i$ curves. A basis on this space is given by $U(1)_i$ networks, which provide the attachment of $U(1)_i$ irreducible representations to each edge $e_i\in\Gamma_i$. The action of fluxes is reproduced as usual in canonical quantization by promoting their Poisson brackets to relations between quantum operators. The vector constraint can be implemented by the analogous procedure adopted in full LQG, {\it i.e.} by summing states over reduced s-knots, which are equivalence class of curves under reduced diffeomorphisms. Also $U(1)_i$ gauge invariance can be realized via standard techniques. In this particular case, invariant intertwiners simply restricts to those states for which at each vertex the representations of the incoming edge and the outcoming edge both along the same direction $i$ are the same. This means that along a given curve $\Gamma_i$ the $U(1)_i$ quantum number is preserved. As a consequence, we cannot realize $U(1)_i$ invariant states along loops and this achievement forbids the definition of the curvature operator, as the trace of the holonomy along a loop, and of the scalar constraint as in Quantum Spin Dynamics (QSD) \cite{qsd}. 

\paragraph{Quantum reduction} The failure of reduced quantization to provide a proper dynamics for the inhomogeneous Bianchi I model forces us to consider a different approach for its quantization. In particular, we propose to reverse the order of reduction and quantization\cite{comp}. 

Hence, let us start from the kinematical Hilbert space of LQG and let us impose a truncation which gives us back the Hilbert space of reduced quantization before imposing the constraints. The kinematical Hilbert space is the space of square integrable $SU(2)$ functionals, whose group elements are attached to piecewise analytic paths, and the kinematical constraints ensure $SU(2)$ gauge invariance and background independence. The first kind of truncation we implement is the one which restricts the edges composing paths to those parallel to fiducial vectors, {\it i.e.} $e_i$ for some $i$. Such a restriction can be realized via a proper projector, which can be shown to project the action of generic diffeomorphisms down to those one of reduced diffeomorphisms. Hence, after this first truncation, the Hilbert space is given by square integrable $SU(2)$ functionals defined over reduced graphs, whose edges are parallel to vectors $\partial_i$, and reduced-diffeomorphisms invariance is the relic of the original background independence.

The emergence of $U(1)_i$ group elements out of $SU(2)$ ones is due to the gauge-fixing procedure associated with the choice of variables (\ref{in}) \cite{io}. Such a gauge-fixing implies that: i) holonomies along $e_i$ belong to the $U(1)$ subgroup generated by $\tau_i$, $U(1)_i$; ii) the following relation holds for fluxes  
\be
\chi_i=\epsilon_{il}^{\phantom{12}k}E_k(S^l)=0,\label{chi}
\ee
$S^l$ being a surfaces dual to $\partial_l$. These two requirements can be realized by mimicking the procedure adopted in Spin Foam models to impose simplicity constraints \cite{eprl}. 
This procedure identifies\cite{ema} the $U(1)_i$ group elements as those ones obtained \emph{by projecting $SU(2)$ ones along the maximum or minimum magnetic components along the direction $i$}. Moreover, the implication of the original SU(2) invariance is \emph{the existence of some reduced intertwiners at vertices}\cite{art,comp}. These intertwiners themselves can give a gauge invariant meaning to closed loops and they allow us to define a sensible dynamics by using the formulation of QSD \cite{qsd} adapted to our reduced model.        
It is worth noting how the volume operator turns out to be diagonal in the $U(1)_i$ network basis, while reduced s-knots must be considered to implement reduced diffeomorphisms invariance. These two features give us \emph{a regularized expression for the scalar constraint, whose matrix elements among basis vectors can be computed analytically}.

\paragraph{Acknowledgments} The work of the author is supported by funds provided by the National Science Center under the agreement DEC-2011/02/A/ST2/00294.

\end{document}